\documentstyle[aps,manuscript,epsf]{revtex}
\begin{document}
\title{\bf Braided Rivers and Superconducting Vortex Avalanches}

\author{Kevin E.\ Bassler$^*$, Maya Paczuski$^{*,\dagger}$, and George F.\ Reiter$^*$}
\address{
$^*$Department of Physics, University of Houston, Houston TX
77204-5506, USA, \\
and $^{\dagger}$Niels Bohr Institute, 17 Blegdamsvej, Copenhagen, Denmark}
\date{\today}

\maketitle 

\begin{abstract}
Magnetic vortices intermittently flow through preferred channels when
they are forced in or out of a superconductor.  We study this behavior
using a cellular model, and find that the vortex flow can make braided
rivers strikingly similar to aerial photographs of braided fluvial
rivers, such as the Brahmaputra.  By developing an analysis technique
suitable for characterizing a self-affine (multi)fractal, the scaling
properties of the braided vortex rivers in the model are compared with
those of braided fluvial rivers.  We suggest that avalanche dynamics
leads to braiding in both cases.
\end{abstract}



\newpage
Magnetic flux penetrates type II superconductors in thin
filaments, or vortices, which can move when an electrical current is
applied.  Since vortex motion creates electrical resistance and
destroys superconductivity, these materials are often produced with
defects that tend to pin vortices.    
Experiments \cite{experiments} and molecular dynamics (MD) simulations
\cite{simulations,braidedvortex,mehta}
have indicated that vortices can move through the pinning landscape
in preferred channels.  Vortices can also move
intermittently in time, or ``avalanche'', as they are forced in or out
of a superconductor \cite{field,behnia}. Similar behavior has also
been seen in simulations
\cite{simulations,braidedvortex,mehta,modelprl}.
Here we use a simple cellular model \cite{modelprl} to mimick the
experiments in Ref.~\cite{field,behnia}, and find that the
intermittent vortex flow can make an intricate braided pattern (Fig. 1),
similar to aerial photographs of braided fluvial rivers, such as the
Brahmaputra, Aichilik, or Hulahula \cite{braidedrivers,sfgres1}.

 Braided rivers \cite{braidedrivers} form a separate class of hydrological systems,
distinct from single-channel rivers and dendritic river  networks.
They are made  of alluvial channels meeting and dividing,
separated by bars and islands.  Their deposits are important
reservoirs of oil, gas, heavy minerals, etc.  Sapozhinikov and
Foufoula-Georgiou (SFG) \cite{sfgres1} have shown that braided rivers
of different scales with different hydrological characteristics
exhibit self-affine  scaling, which may be universal.

Here we extend the SFG method and show that  braided rivers in
the vortex model exhibit self-affine multifractal behavior over two
decades in length scale.  We compare our results with the correlation
integral method of SFG, which measures the scale dependence of one
particular moment of the flow, and propose that our
multifractal analysis technique, which probes all moments, be used to
further characterize braided fluvial rivers as well as other braided
systems.  We suggest that the stick-slip dynamics of vortices leading
to avalanches of magnetic flux in superconductors may be analogous to
pulse load transport \cite{gomez}  of avalanching  sediment in fluvial systems, and lead to
braiding in both cases, despite their vastly different length scales and
microscopic descriptions.  Comparing our measurements with those
of SFG (where possible) leads us to speculate that these braided
systems could represent a single universality class of dynamic
critical phenomena.

The cellular model \cite{modelprl} describes the motion of driven
vortices at large physical length scales, and includes basic features
of vortex dynamics: over-damped motion of vortices, repulsive
interactions between vortices, and attractive pinning interactions at
defects in the material.  As in experiments, vortices are slowly
pushed into the system at one boundary  and allowed to leave
at the other boundary.  Within a range of parameters, the
model evolves into a self-organized critical state \cite{soc} with avalanches of
all sizes, as explained in detail in \cite{modelprl}.  Self-organized
criticality has also been observed experimentally \cite{field,behnia} 
for a limited range of temperatures and magnetic
fields, and in MD simulations of the microscopic
equations of motion \cite{olson}.

We briefly summarize the model.  Consider a
two-dimensional honeycomb lattice where each cell $x$ has three
nearest neighbors, and is occupied by an integer number of vortices
$m(x)$.  The force to
move a vortex from $x$ to $y$ is 
\begin{eqnarray*}
F_{x \rightarrow y} & = & V_{\mbox{pin}}(y) - V_{\mbox{pin}}(x) + [m(x) - m(y) -1] \\
& & + r[m(x1) + m(x2) - m(y1) - m(y2)] \quad .
\end{eqnarray*}
 The nearest
neighbor cells of $x$ are $y$, $x1$, and $x2$, and the nearest
neighbors cells of $y$ are $x$, $y1$, and $y2$, and $r<1$.
The  pinning
potential $V_{\mbox{pin}}(x)$  is equal to
to $V_{max}$ with probability $p$ and $0$ with probability $1-p$.
The numerical results
presented here are for $r=0.1$, 
$V_{max}=2.0$, and $p=0.1$. However, the results are robust over a limited range of
the parameters, as discussed below. 
In one iteration, at each cell
a single vortex moves one lattice unit when the force in that direction is
positive. If more than one unstable direction exists, one of them is
chosen randomly \cite{annealed}.
 A vortex  reaching the right edge of the system is
removed. Periodic boundary conditions are applied at the top and
bottom of the lattice.   An avalanche is initiated by adding a vortex to a stable
configuration on the left edge of the system. It proceeds by
repeatedly updating  until the configuration is again
stable.  Then a new vortex is added to the system.

The spatial variation of the flow is measured in terms
of the number of vortices moving in each cell, averaged over a very long
time interval representing many vortices flowing through the system.
Fig.~1.  represents a ``time-lapsed'' photograph of vortex motion.
Rather than exhibiting uniform flow, the vortices clearly have
preferred channels to move in.   In
sufficiently large systems,  the channels form an
intricate braided pattern.  In fact braiding was observed, at 
smaller length scales, from MD simulations 
\cite{braidedvortex,mehta}, for a range of parameter
values. Outside of this range, branched, ``Hortonian'' structures
resembling dendritic river networks were observed.  Nevertheless, a
direct comparison, using techniques developed to characterize braided
systems \cite{sfgres1}, with
braided fluvial rivers \cite{braidedrivers} has not previously been
made.

 For a given configuration of pinning centers in our model, the
observed braided pattern is absolutely stable, independent of initial
conditions, including a system empty of vortices, and a system that initially
had a uniform average slope greater than the critical slope.  However,
the pattern changes if the pinning centers are moved.  Thus, in some
superconductors maintained at sufficiently low temperatures, we predict
that a braided river pattern would exist that would be stable and
reproducible for a given sample, if one averaged the flow over a long
time interval. In fact, Battacharya and Higgins \cite{fingerprints}
have reported sample dependent ``fingerprints'' in IV measurements
which is likely an experimental verification of this behavior.

Qualitatively, the braided vortex river pattern we observe resembles patterns of
interconnected channels formed by water flowing over non-cohesive
sediment.  In fact, braiding has
been proposed to be the fundamental instability of laterally
unconstrained free surface flow over cohesionless beds, and has been
found to be a robust feature in simulations of river flow with
sediment transport that includes both erosion and redeposition
\cite{naturepaper}.   Interestingly, where no redeposition occurs,
 as in mountainous regions with high slope, branched, dendritic
networks are obtained.  

In order to examine the possible relationship between these two types
of braided systems, we make a quantitative analysis of braided vortex
rivers, and compare the results with those previously measured from
aerial photographs and radar imagery \cite{sfgres1} of different
fluvial rivers.  Although the pinning sites are randomly distributed
in our model, with no extended spatial correlations, the
braided structure itself shows long-range correlations.  Since each
individual flow event, or avalanche, has no characteristic scale, the
sum of those events, as revealed by the pattern of flow in Fig. 1,
also is scale-free up to the size of the system.

A braided river  can be characterized by its behavior under a change
of scale.  The braided vortex river
is a self-affine multifractal over the range of length scales we are able
to study.  The
self-affinity is due to the fact that the average flow is anisotropic 
since it follows the
average slope of the vortex pile.  This makes the direction of flow and
perpendicular to flow distinct.
In order to reproduce the
same structure under a change of scale one has to scale the two
directions in a different manner.  This effect is also seen in
braided fluvial rivers, as sediment flows down an average
slope due to gravity.  The multifractality reflects
spatial intermittency.  Intermittency here means that almost all cells in the model 
contain at least a small portion of the vortex flow, but
the flow is highly concentrated into a subset of the cells, with the most
highly concentrated flow, dominating the highest moments,
 occurring in filamentary structures indicated
in yellow in Fig. 1.  If each moment of the flow pattern exhibits scaling with a different
dimension  then the flow is multifractal.

A standard mathematical characterization of the scaling properties of
braided rivers is made
by partitioning the pattern, such as Fig. 1, into rectangular cells,
each one of size $X \times Y$ elementary cells. 
If $P_i$ is defined to be the
fraction of the overall flow that falls into $i$th rectangle of
the partitioning, then the qth moment of the probability partition
is
\begin{equation}
M_q(X,Y) = \sum_{i} P_i^q \quad {\rm for}
\quad q \geq 0 \,\, ,
\label{moment}
\end{equation}
where the sum is over all rectangles of size $X \times Y$
needed to cover the entire pattern.  
The spectrum of fractal dimensions $\{ D_q \}$ can be calculated by
determining the scaling behavior of $M_q(X,Y)$
when $X \sim Y \sim L$ by
$$
M_q(L)  \sim L^{-(1-q) D_q} .
$$
Because of the singular nature of this definition, special care must
be used to define $M_q$ for $q=0$ and $q=1$ \cite{grassberger}.  The
exponents $D_0$, $D_1$, and $D_2$ correspond to the capacity,
information, and correlation dimensions, respectively
\cite{grassberger}.  The results of our analysis of the braided vortex
rivers are shown in Fig.~2.  For the largest system size we could
study, each moment, $M_q$, exhibits scaling with a well-defined
dimension, $D_q$, over two decades in length scale.  The measured
dimensions vary continuously from $D_0 = 2.0$ to $D_8 \approx 1.4$;
such a variation indicates that the pattern is ``multi-fractal''.  The
value of $D_0$ is to be expected since virtually all of the elementary
cells contain a non-zero portion of flow.  The lower values of $D_q$
for the higher moments, $q$, arise from the filamentary structure of
highly concentrated activity, as mentioned above.  The dimension $D_2$
has been measured for braided fluvial rivers with a variety of
different types of sediment beds over a wide range of length scales,
obtaining $D_2=1.5-1.7$ \cite{braidedrivers,sfgres1}.  This range is
relatively close to the values ($D_2 \sim 1.8$) we have
measured for different realizations of our model.  As discussed later,
our accuracy is not sufficient to ascertain whether the small apparent
difference is significant or not.  A multifractal analyses of braided
fluvial rivers using the method described here could determine if
they are multifractal or not.

In order to establish and characterize the anisotropic scaling
 properties of the braided pattern, we have adapted and extended the method of
 SFG
\cite{sfgres1,sfgmethod,difference}, which was
 developed to characterize the self-affine structure of
  braided fluvial rivers.
 We extend their method to take into account the fact that the
 self-affine structure can also be multifractal, as demonstrated
above.  The most general
 scaling of a pattern is found by determining the scaling
 of $M_q(X,Y)$ when the rectangle dimension $X \times Y$ scales such
 that $X \sim Y^{\nu_x/\nu_y}$.  In this case, the scaling hypothesis
 is
\begin{equation}
{M_q(X_1,Y_1)\over M_q(X_2,Y_2)} 
= \left( {X_1 \over X_2} \right)^{1/(q-1) \; \nu_x(q)}
= \left( {Y_1 \over Y_2} \right)^{1/(q-1) \; \nu_y(q)}.
\label{scaling}
\end{equation}

Defining $z_q = \log M_q$, $x = \log X$, and $y = \log Y$, and following
the arguments of Ref. \cite{sfgmethod}, for each $q$, we find
\begin{equation}
(q-1) \; \nu_x(q) \; {d z_q \over d x} + (q-1) \; \nu_y(q) \; {d z_q \over d y} = 1
\label{cylindereqn}
\end{equation}
Taking the logarithmic derivatives of the moment of the probability
partition $dz_q / dx$ and $dz_q / dy$ numerically at a number of
different points $(x,y)$, the results can be the fit into
(\ref{cylindereqn}) to obtain $\nu_x(q)$ and $\nu_y(q)$. The results
of this analysis on the braided vortex river patterns are shown in
Fig.~3.  Note that the exponents $\nu_x$ and $\nu_y$ vary with $q$.
This variation is consistent with the variation of $D_q$ described
above.  In simple terms, it means that the different fractals which
are formed by different moments of the flow also exhibit different
self-affine properties, with the highest moments, which are dominated
by long, filamentary objects, being more anisotropic than the lower
moments.

Previous analyses on different braided fluvial rivers have determined
$\nu_x(2) = 0.72-0.77$ and $\nu_y(2) = 0.47-0.52$ \cite{sfgres1}.
From the results in Fig.~3, we obtain $\nu_x(2) = 0.67 \pm 0.03$ and
$\nu_y(2) = 0.44 \pm 0.03$ for braided vortex rivers, where the error
bars only represent statistical errors based on the data shown.
However, there are also potentially significant sources of systematic
error due to finite size limitations.  Noticing boundary effects
apparent in Fig.~1, we analyze the scaling only in a strip in the
center, typically beginning and ending 80 lattice cells inside the
left and right edges.  We have varied both the strip width and the
numerical differentiation scheme and observed the variation in the
apparent scaling dimensions.  We estimate the systematic error to be
approximately 10\% of the quoted value for all exponents.  Within
error, including systematic error, our results for $\nu_x(2)$ and
$\nu_y(2)$ are consistent with those measured for different braided
fluvial rivers, although we cannot rule out that the exponents are
different.  The anisotropic exponents $\nu_x$ and $\nu_y$ for other
values of $q$, have not  been measured for braided fluvial
rivers, which would be necessary to make firmer statements about the
quantitative similarities or differences between these two systems.
Nevertheless, since the three exponents where we can compare do not
exhibit differences outside the error limits, we are lead to speculate
that the braided structures in the two systems may represent a single
universality class.

In this regard, it is important to note
that we have also varied the parameters of our model and studied the resulting
river structure. Over a range of parameters, $r \in [0.1, 0.2]$, 
$V_{max} \in [1.0, 5.0]$, and $p \in [0.1, 0.4]$, we observe a similar braided 
pattern with fractal and multi-fractal exponents that lie within the  error
bounds of the results reported here. However, sufficiently
outside this range of parameters 
it is likely that the morphology of the river structure is different. For example,
at weaker pinning strengths the rivers become broad and the fractal structure 
disappears. At stronger pinning strengths,
the braided structure also seems to disappear
and the pattern formed by the rivers more closely resembles dendritic river networks.
These findings concerning the different river morphologies are similar to 
those reported for MD simulations \cite{mehta}. However, 
further study is required to confirm these results. 
  
It has previously been postulated
that braiding of fluvial rivers is due to a self-organized critical
process \cite{sfgsoc1}, but a mechanism for this to occur has not been
identified.  
Are there avalanches in fluvial rivers that could lead to self-organization and
produce the observed braiding?  In fact there are.  ``Pulses'' in
bedload transport have been observed to occur on all spatial and
temporal scales up to those limited by the size of the river studied
\cite{gomez}.  We have observed analogous pulses in our model by
measuring the vortex flow through individual lattice cells as a
function of time.  Perhaps vortices of magnetic flux are analogous to sediment in
fluvial rivers.  The elementary stick-slip process is that of sediment
slipping and then resticking at some other point \cite{naturepaper},
like our intermittently moving vortices.  The elementary slip event
can dislodge nearby sediment leading to a chain reaction of slip
events, or avalanches.  Sediment transport can be triggered when the
local sediment slope is too high; the same is true for vortices in a
superconductor.  Thus, in both magnetic flux and fluvial rivers it
appears that the braiding could emerge through a stick-slip process
consisting of avalanches of all sizes.

In should also be noted that although there is disorder present 
in fluvial rivers, due to inhomogeneities in the bedload, it is
different than the disorder due to pinning in superconductors. 
The location of the pinning centers  is quenched, whereas
the location of bedload inhomogeneities in fluvial rivers are annealed at some time scale.
Thus the braided patterns produced
in our simulations are fixed, while the braided patterns observed in
fluvial rivers are not \cite{sfgsoc1}.  Nevertheless,
despite this difference, the resulting braided patterns are similar.

K. B. and G. R. thank the Texas Center for
Superconductivity at the University of Houston
for partial support of their research. We thank A. Rinaldo for pointing
out references.

\begin{figure}
\narrowtext
\caption{
 {\bf Braided Rivers of Superconducting Vortices.} A
``time-lapsed'' photograph of vortex motion with average flow from
left to right.  The lattice size is $600 \times 500$.  Sites
containing an average amount of flow are shown in red.  Yellow sites
have a flow level greater than 20 times the average.  Dark blue sites
have almost no vortex flow, although virtually every site has some
minimal amount.  The intricate braiding pattern is remarkably similar
to the pattern formed by braided fluvial rivers.}
 \end{figure}

\begin{figure}
\narrowtext
\caption{
 {\bf Multi-fractal scaling exponents.} The scaling of moments
 as a function of $L$, calculated from
Fig.~1. The slopes of the lines are the exponents
 $D_q$. Shown here, from
 bottom to top, are results for $q =$ 0, 1, 2, 3, 4, 6. Inset shows
 the dimensions $D_q$ as a function of $q$ for five different
 realizations of braided rivers resulting from the same parameters,
 with different locations for the pins.}
 \end{figure}

\begin{figure}
\narrowtext
\caption{
{\bf Self-affine multi-fractal exponents.} Results for the
anisotropic multi-scaling exponents $\nu_x$ and $\nu_y$ as a function
of $q$ for five different braided river patterns. 
Results were calculated from
the same data sets used in the inset in Fig.~2. The dark
filled symbols correspond to the results for $\nu_x$, while open symbols
correspond to the results for $\nu_y$. The results for each of the
five different patterns are indicated with unique symbols. 
The average over the five runs is indicated by the circles connected by the 
thick line.}
 \end{figure}


\begin{references}
\bibitem{difference} Sapozhiknov and Foufoula-Georgiou analyzed the
self-affine scaling of the correlation integral where as we analyize
the self-affine scaling of moments of the probability
partition. However, the scaling of the second moment of the
probability partition ($q=2$) is equivalent to the scaling of the
correlation integral \cite{grassberger}.
\end{references}
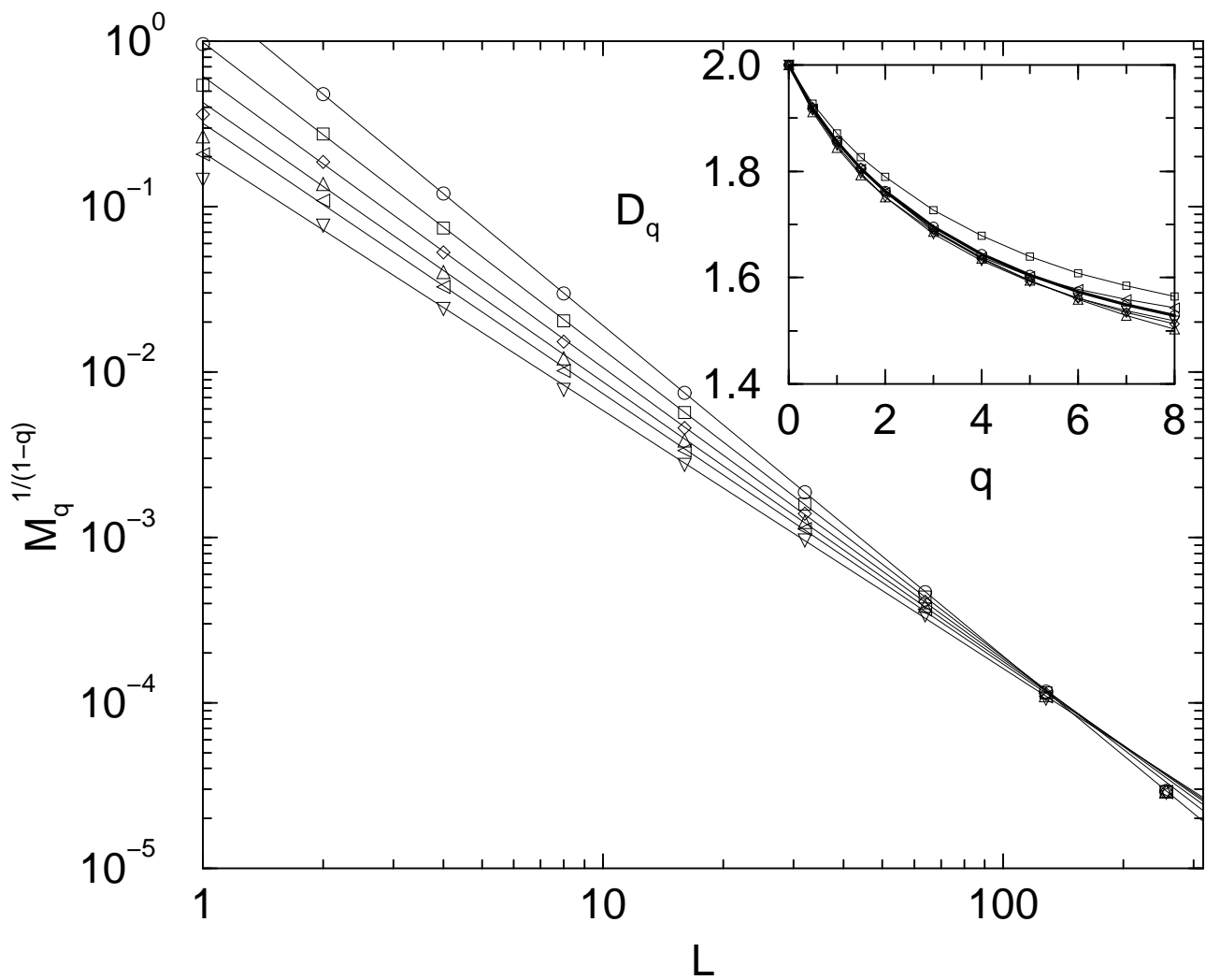
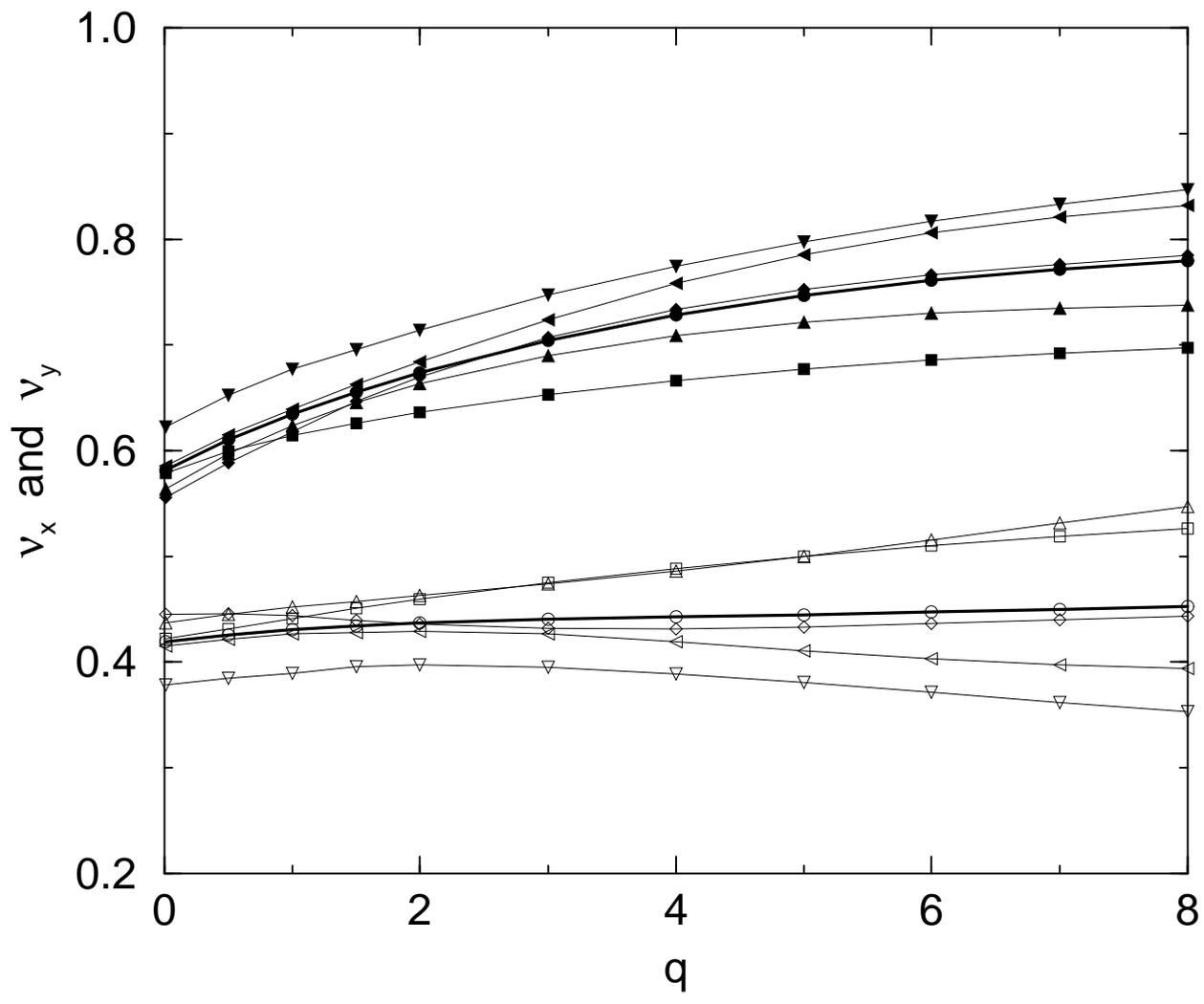
\end{document}